\documentclass[aps,pra,floatfix,twocolumn,english,showpacs]{revtex4}
\usepackage[latin1]{inputenc}
\usepackage{float}
\usepackage{amsmath}
\usepackage{graphicx}

\makeatletter

\providecommand{\tabularnewline}{\\}

\usepackage{babel}
\makeatother
\begin{document}

\title{Quantum State Tomography: `the best' is the enemy of `good enough'.}

\author{Max S. Kaznady and Daniel F. V. James}

\email{max.kaznady@gmail.com}

\email{dfvj@physics.utoronto.ca}

\affiliation{Department of Physics and Centre of Quantum Information and Quantum
Control\\
University of Toronto, 60 St. George Street, Toronto, ON,  M5S 1A7, CANADA}

\date{\today}

\begin{abstract}
In this paper, we examine a variety of strategies for numerical quantum-state estimation from
data of the sort commonly measured in experiments involving quantum state tomography.
We find that, in some important circumstances, an elaborate and time-consuming numerical
optimization to obtain  `the best' density matrix corresponding to a given data set is not necessary,
and that cruder, faster numerical techniques may well be `good enough'.

\end{abstract}


\maketitle

%%%%%%%%%%%%%%%%%%%%%%%%%%%%%%%%%%%%%%%%%%%%%%%%%%%%%%%%%%%%%%%%
%%%%%%%%%%%%%%%%%%%%%%%%%%%%%%%%%%%%%%%%%%%%%%%%%%%%%%%%%%%%%%%%
%%%%%%%%%%%%%%%%%%%%%%%%%%%%%%%%%%%%%%%%%%%%%%%%%%%%%%%%%%%%%%%%
\section{Introduction}
%%%%%%%%%%%%%%%%%%%%%%%%%%%%%%%%%%%%%%%%%%%%%%%%%%%%%%%%%%%%%%%%
%%%%%%%%%%%%%%%%%%%%%%%%%%%%%%%%%%%%%%%%%%%%%%%%%%%%%%%%%%%%%%%%
%%%%%%%%%%%%%%%%%%%%%%%%%%%%%%%%%%%%%%%%%%%%%%%%%%%%%%%%%%%%%%%%
The goal of quantum state tomography~\cite{leonhardt, ParisRehacek, Hayashi}
is to estimate, from a series of projective measurements performed on identically
prepared quantum systems, the density matrix of the underlying ensemble
of which these quantum systems are realizations. This process is necessarily
non-deterministic in nature, relying on the frequency of experimental
outcomes to estimate probabilities - a process that converges to the
actual probabilities only in the infinite limit. Thus the reconstruction
of the quantum state cannot be exact in any realistic experiment.
Furthermore, these measurements can only yield estimates of the
on-diagonal elements of the density matrix, but not directly
any data about the off-diagonal elements. It is
necessary to perform various unitary operations on the system (or,
equivalently to perform projective measurements in a variety of bases)
in order to obtain such information about the complete state. Indeed,
for a system with a discrete spectrum of $n$-levels, the density
matrix is specified by $n^{2}-1$ independent real parameters, and each parameter
will require a separate measurement.  Even after the required measurements
have been performed, the experimenter
faces the problem of estimating the density matrix from incomplete and
noisy data. The problem is aggravated by the constraints that quantum
physics places on the density matrix: It must be a non-negative, unit-trace Hermitian matrix.
Today, the approach that is usually taken is to determine computationally
what is the `best' such positive, unit-trace Hermitian matrix which
corresponds to a particular data set, and what confidence can we place
on such an estimate. The most complicated such tomographic measurement
performed to date~\cite{Haffner}, on an 8 qubit (256 state) system,
realized in a trapped ion experiment, was {\em limited not by the experimental
capabilities of the system, but rather by the complexity of the numerical state
recovery problem}~\cite{Haffnerpc}. This computational complexity, while underscoring
the awesome computational potential inherent in quantum information, nevertheless
presents an experimenter, intent on exploring larger and larger Hilbert spaces, with considerable tribulation when characterising the performance of his or her apparatus.

In this paper we examine the problem from an entirely computational
perspective. Specifically, we address the concern that maybe we are
being too fastidious in approaching the state reconstruction problem.
One can obtain a positive, unit-trace Hermitian matrix from tomographic
data in a variety of ways. First, and most simply, one could generate
a linear reconstruction of the noisy data (which tends to give a non-positive
matrix), and ensure positivity by setting the negative eigenvalues
to zero, then re-normalizing to ensure a unit-trace. This we call the ``Quick and Dirty" (QD) approach.
A second strategy is to assume the state must be nearly pure
- after all, quantum technologies are usually in the business of trying to create pure
states - and to simplify the computation by finding the pure state most compatible with the data.
We call this the ``Forced Purity'' (FP) approach. A third approach is full optimization, i.e.
the application of some constrained optimization routine, with a specific metric
to define the `distance' between our data set and a positive
density matrix, and search parameter space until the absolute `best' (i.e. global minimum)
density matrix is obtained. Our goal is specifically to address the
question: When is the rigorous optimization required, and when will
some short-cut technique be good enough? This is a question that can
only be addressed by simulation: Since we need to know \emph{a priori}
the underlying density matrix of the ensemble to compare the recovered
estimates. Starting with assumed density matrix, we employ a pseudo-random
number generator to create some `pseudo-experimental data' with
appropriate probability distribution. The various approaches to density
matrix recovery are applied to it, and the result
is compared with the initial density matrix to assess the accuracy of the recovery techniques.
Our analysis concerns solely multiple correlated two-level systems, e.g. the qubits of a small
scale quantum computer; however, many of the techniques and results we present are
readily adaptable to more general systems.

The paper is organized as follows: In Sec.~\ref{sec:One-qubit}
we discuss the generic tomography problem for a single qubit, which
is generalized to the n-qubit case in Sec.~\ref{sec:Generalization-to-N-qubits},
describing specifically a number of memory management techniques required
for scalability of the code, and our novel approach to the optimization
routine (using gradient-based algorithms and employing the matrix
differential calculus). The code itself is described in detail in
Sec.~\ref{sec:Description-of-Code}, and our results in Sec.~\ref{sec:Results}.

%%%%%%%%%%%%%%%%%%%%%%%%%%%%%%%%%%%%%%%%%%%%%%%%%%%%%%%%%%%%%%%%
%%%%%%%%%%%%%%%%%%%%%%%%%%%%%%%%%%%%%%%%%%%%%%%%%%%%%%%%%%%%%%%%
%%%%%%%%%%%%%%%%%%%%%%%%%%%%%%%%%%%%%%%%%%%%%%%%%%%%%%%%%%%%%%%%
\section{One qubit \label{sec:One-qubit}}
%%%%%%%%%%%%%%%%%%%%%%%%%%%%%%%%%%%%%%%%%%%%%%%%%%%%%%%%%%%%%%%%
%%%%%%%%%%%%%%%%%%%%%%%%%%%%%%%%%%%%%%%%%%%%%%%%%%%%%%%%%%%%%%%%
%%%%%%%%%%%%%%%%%%%%%%%%%%%%%%%%%%%%%%%%%%%%%%%%%%%%%%%%%%%%%%%%
In this section, we will review the basic concept of quantum state
tomography by considering the estimation of a state of a single two-level system, or qubit.

%%%%%%%%%%%%%%%%%%%%%%%%%%%%%%%%
\subsection{Parametrizing the Density Matrix \label{sub:Density-Matrix}}
%%%%%%%%%%%%%%%%%%%%%%%%%%%%%%%%
The density operator describing the state of a system~\cite{CT} is a Hermitian,
non-negative definite operator of unit-trace.
The set of Pauli matrices~\cite{CT2} $\{\hat{\sigma}_{0},\hat{\sigma}_{1}{,\hat{\sigma}}_{2}{,\hat{\sigma}}_{3}\}$
form, for a two dimensional space, a complete orthonormal set of matrices,
%with respect to the Hilbert-Schmidt inner product and,
so that $\hat{\rho}$ can be expanded as a linear combination
of $\hat{\sigma}_{\mu}$ as
\begin{equation}
\hat{\rho}=\sum_{\mu=0}^{3}r_{\mu}\hat{\sigma}_{\mu},
\label{rho_from_r}
\end{equation}
where
\begin{equation}
r_{\nu}=Tr\{\hat{\sigma}_{\nu}\hat{\rho}\}/2.
\end{equation}
Since $Tr\{\hat{\rho}\}=1$, $r_{0}=1/2$; further, since $\hat{\rho}^{\dagger}=\hat{\rho}$, the $r_{\nu}$ are
all real parameters.

The $r_{\nu}$ may be determined experimentally as follows:
Suppose we perform a measurement, specified by the projector $\hat{\Pi}_0$, on
the system, the probability of obtaining a positive outcome is $Tr\{\hat{\rho}\hat{\Pi}_0\}$.
Repeating this measurement ${\cal N}$ times on identically prepared systems, the expected
number of times we obtain this outcome will be
\begin{eqnarray}
n_{0}&=&{\cal N}Tr\{\hat{\rho}\hat{\Pi}_0\}\nonumber\\
&=&{\cal N}\sum_{\mu=0}^{3} Tr\{ \hat{\sigma}_{\mu} \hat{\Pi}_0\}  r_{\mu}.
\label{eq:n sub nu from density matrix}
\end{eqnarray}
If one repeated this procedure of multiple measurements for a set of four different measurement operators, $\{ \hat{\Pi}_{\nu}\}, (\nu=0,1,2,3)$
one obtains a set of linear equations
\begin{equation}
n_{\nu}={\cal N}\sum_{\mu=0}^{3}B_{\nu,\mu} r_{\mu},
\end{equation}
where
\begin{equation}
B_{\nu,\mu}=Tr\{ \hat{\sigma}_{\mu} \hat{\Pi}_{\nu}\}.
\label{mightymouse}
\end{equation}
By choosing the measurement operators, $\{ \hat{\Pi}_{\nu}\}$, judiciously, one can ensure that $B_{\nu,\mu}$ is non-singular,
and hence that the desired parameters $r_{\mu}$ can be obtained from the observed quantities
$n_{\nu}$, viz.,
\begin{equation}
r_{\nu}=({\cal N})^{-1}\sum_{\mu=0}^{3}(B^{-1})_{\nu,\mu}n_{\mu}.\label{eq:Inverted-B}
\end{equation}
Substituting $r_{\nu}$ into Eq.~\eqref{rho_from_r}, we obtain the density matrix, as a function of measurement
outcomes, {\em provided the measurements have no noise or errors in them}.

Following the precedent of Ref.~\cite{DFVJ Measurement of qubits}
we use the standard Stokes measurement basis for our numerical
experiments. These measurement operators are given by:
\begin{equation}
\begin{array}{cc}
\hat{\Pi}_{0}=\frac{1}{2}(|0\rangle\langle 0|+|1\rangle\langle 1|), & \hat{\Pi}_{1}=|0\rangle\langle 0|,\\
\hat{\Pi}_{2}=|\bar{D}\rangle\langle\bar{D}|, & \hat{\Pi}_{3}=|R\rangle\langle R|,\end{array}\label{eq:Stokes-Basis}\end{equation}
where $|0\rangle$ and $|1\rangle$ represent the two states of our qubits, and
\begin{eqnarray}
|R\rangle&=&\frac{1}{\sqrt{2}}(|0\rangle-i\cdot|1\rangle),\\
&&\nonumber\\
|\bar{D}\rangle&=&\frac{1}{\sqrt{2}}(|0\rangle-|1\rangle).
\end{eqnarray}

A natural metric to compare the recovered density matrix $\hat{\rho}_\text{meas}$
with the actual density matrix $\hat{\rho}_\mathrm{true}$ is the {\em fidelity}~\cite{Jozsa},  defined as:
\begin{equation}
F(\hat{\rho}_\mathrm{meas},\hat{\rho}_\mathrm{true})=
\{ Tr[(\sqrt{\hat{\rho}_\mathrm{meas}}\hat{\rho}_\mathrm{true}\sqrt{\hat{\rho}_\mathrm{meas}})^{1/2}]\}^{2}.\label{eq:Fidelity}
\end{equation}
However, when we invert the measurement data linearly, our recovered ``density matrix'' $\hat{\rho}_\mathrm{linear}$ is not non-negative definite and hence
we have the specific problem that fidelity turns out to be complex (not to mention the more general problem that
$\hat{\rho}_\mathrm{linear}$ cannot be interpreted as a density matrix of a physical state).
We have to correct the matrix obtained by linear reconstruction to obtain a proper density matrix.

%%%%%%%%%%%%%%%%%%%%%%%%%%%%%%%%
\subsection{{}``Quick and Dirty'' Reconstruction}
%%%%%%%%%%%%%%%%%%%%%%%%%%%%%%%%

As a simple initial approach to this problem, we can decompose $\hat{\rho}_\mathrm{linear}$ into its spectral representation, i.e.
\begin{equation}
\hat{\rho}_\mathrm{linear}=\hat{U}\hat{D}\hat{U}^{\dagger},
\end{equation}
where $\hat{D}$ is the diagonal matrix of eigenvalues (which are real, but not necessarily positive) and $\hat{U}$
is a unitary matrix. We then set all negative eigenvalues in $\hat{D}$
to zero, call this matrix $\hat{D}^{\prime}$, and obtain:
\[
\hat{\rho}_\mathrm{QD}=\frac{\hat{U}\hat{D}^{\prime}\hat{U}^{\dagger}}{Tr\{\hat{D}^{\prime}\}}.
\]
This provides a rough initial estimate of the state; one of the goals of our analysis in this paper is to assess how good an estimate it is.

%%%%%%%%%%%%%%%%%%%%%%%%%%%%%%%%
\subsection{``Forced Purity"\label{subsection:FP}}
%%%%%%%%%%%%%%%%%%%%%%%%%%%%%%%%
An alternative approach to the problem of obtaining a non-negative definite density matrix from measured data is to assume that the state is {\em pure}.
Recall that for a pure state $|\Psi\rangle$ the density matrix can be described by a single ket as
$\hat{\rho}_\mathrm{pure}=|\Psi\rangle\langle\Psi|$. Such a density matrix for $n$ qubits has eigenvalue $0$
with degeneracy $2^{n}-1$ and eigenvalue $1$ with degeneracy $1$.

Because $\hat{\rho}_\mathrm{pure}$ is also Hermitian, it can be written
in its spectral decomposition as
\[
\hat{\rho}_\mathrm{pure}=\hat{V}\hat{D}_\mathrm{pure}\hat{V}^{\dagger},
\]
where $\hat{D}_\mathrm{pure}$ is the diagonal matrix with a single element equal to $1$,
and all other elements being zero; $\hat{V}$ is a unitary matrix.

During linear inversion of a pure state, the eigenvalues of $\hat{\rho}_\mathrm{linear}$
may be negative, but sufficiently close to eigenvalues of $\hat{\rho}_\mathrm{pure}$.
The idea of forcing purity on such a state is to obtain
\[
\hat{\rho}_\mathrm{FP}=\frac{\hat{V}_\mathrm{linear}\hat{D''}\hat{V}_\mathrm{linear}^{\dagger}}{Tr\{\hat{D''}\}},
\]
where $\hat{D''}$ is the diagonal matrix obtained from  $\hat{D}$
by setting the largest eigenvalue equal to $1$, and all others equal to $0$.

%%%%%%%%%%%%%%%%%%%%%%%%%%%%%%%%
\subsection{Maximum Likelihood}
%%%%%%%%%%%%%%%%%%%%%%%%%%%%%%%%
Any Hermitian $2\times2$ non-negative unit-trace matrix can be
uniquely parametrized using the Cholesky decomposition as:

\begin{equation}
\hat{\rho}_\mathrm{ideal}(t_{1},\, t_{2},\, t_{3},\, t_{4})=\frac{T^{\dagger}T}{Tr\{ T^{\dagger}T\}},
\end{equation}
where
\begin{equation}
T(t_{1},\, t_{2},\, t_{3},\, t_{4})=
\left(\begin{array}{cc}
    t_{1} & 0\\
    t_{3}+i\cdot t_{4} & t_{2}
    \end{array}\right).
\end{equation}
Thus a `physical' density matrix can be specified by the four parameters $\vec{t}=\{t_1,t_2,t_3,t_4\}$.  The ideal
of the maximum likelihood method is to perform a search of the $\vec{t}$ parameter space until we find
a $\hat{\rho}_\mathrm{ideal}\left(\vec{t}\right)$ which is most likely to have generated the observed data $\{n_0,n_1,n_2,n_3\}$.
To assess this likelihood, suppose that each datum $n_{\mu}$ is a statistically independent, Poisson-distributed random variable with
expectation value  $\bar{n}_{\mu}$.  Further, if $\bar{n}_{\mu}$ is a large number, the Poisson distribution is well approximated by the
Gaussian distribution, i.e.
\begin{equation}
P(n_0,n_1,n_2,n_3)=\frac{1}{N_\mathrm{norm}}\prod_{\nu=0}^{3}\exp\left[-\frac{(n_{\nu}-\bar{n}_{\nu})^{2}}{2\bar{n}_{\nu}}\right],
\label{fredflinstone}
\end{equation}
where $N_\mathrm{norm}$ is the normalization constant.
If each datum $n_{\mu}$ is garnered from ${\cal N}$ repetitions of a measurement carried out on a system in state $\hat{\rho}_\mathrm{ideal}\left(\vec{t}\right)$, it is
reasonable to make the identification $\bar{n}_{\nu}(t_{1},t_{2},t_{3},t_{4})={\cal N}\langle\psi_{\nu}|\hat{\rho}_\mathrm{ideal}(t_{1},t_{2},t_{3},t_{4})|\psi_{\nu}\rangle$, and the
likelihood of a given parameter vector $\vec{t}$ generating the data $\{n_0,n_1,n_2,n_3\}$ can be obtained by substituting this identity into Eq.~\eqref{fredflinstone}.
We are then in a position to determine the parameter vector for which this probability is maximized, and hence the most likely density matrix.  Instead of maximizing  Eq.~\eqref{fredflinstone}, it is equivalent, and mathematically more convenient to minimize the following function:
\begin{equation}
{\cal L}(\vec{t})=\frac{1}{2}\sum_{\nu=0}^{3}
\frac{
    \left[n_{\nu}-{\cal N}
    Tr\{\hat{\Pi}_{\nu}\hat{\rho}_\mathrm{ideal}(\vec{t})\}
    %\langle\psi_{\nu}|\hat{\rho}_\mathrm{ideal}(\vec{t})|\psi_{\nu}\rangle
    \right]^{2}
        }{
    {\cal N}Tr\{\hat{\Pi}_{\nu}\hat{\rho}_\mathrm{ideal}(\vec{t})\}
        }.
\label{eq:L-fn}
\end{equation}

In order to optimize this function efficiently, we need to compute
its gradient. This is not an easy feat, as the closed analytic form
does not simplify well, and finite-differencing is too inefficient.
The situation becomes exponentially worse as we increase the number
of qubits.

%%%%%%%%%%%%%%%%%%%%%%%%%%%%%%%%%%%%%%%%%%%%%%%%%%%%%%%%%%%%%%%%
%%%%%%%%%%%%%%%%%%%%%%%%%%%%%%%%%%%%%%%%%%%%%%%%%%%%%%%%%%%%%%%%
%%%%%%%%%%%%%%%%%%%%%%%%%%%%%%%%%%%%%%%%%%%%%%%%%%%%%%%%%%%%%%%%
\section{Generalization to N-qubits \label{sec:Generalization-to-N-qubits}}
%%%%%%%%%%%%%%%%%%%%%%%%%%%%%%%%%%%%%%%%%%%%%%%%%%%%%%%%%%%%%%%%
%%%%%%%%%%%%%%%%%%%%%%%%%%%%%%%%%%%%%%%%%%%%%%%%%%%%%%%%%%%%%%%%
%%%%%%%%%%%%%%%%%%%%%%%%%%%%%%%%%%%%%%%%%%%%%%%%%%%%%%%%%%%%%%%%
In the previous section, we outlined the possible routines for performing
tomography of a single qubit. We now extend these routines to
a higher number of qubits and see how the ``Quick and Dirty" and ``Forced
Purity" methods compare to the elaborate and time-consuming Maximum Likelihood Estimation (MLE) routine.

At first, the problem looks very simple - any state of each qubit
is completely characterized by only 4 measurements. Hence, numerically the MLE procedure is rather
easy to implement - we just need to optimize a function of 4 variables,
which is achieved by the simplex or Powell optimization algorithm in
a fairly short amount of time~\cite{Numerical Recipes}, without
computing the gradient. However, two qubits,
when correlated, are not characterized by 8 measurements, but by $4\cdot4=16$
measurements, because we are looking at a \emph{system} of 2 qubits.
If \emph{n} is the number of qubits, then we would need to obtain
$4^{n}$ measurement outcomes in some fixed $4^{n}$ dimensional basis.
Due to wave function collapse, we can only perform one projection measurement at a time (an outcome is an average over multiple identical projection measurements) and for each projection measurement on one qubit, we have to
cycle
through all possible combinations of projection measurements for the
other qubits.

Let us introduce the following set of operators which generalize the Pauli matrices for $n$ qubit systems:
\begin{equation}
\hat{\Gamma}_{\mu}=\frac{1}{\sqrt{2^n}}\hat{\sigma}_{\mu_{1}}\otimes\hat{\sigma}_{\mu_{2}}\otimes...\otimes\hat{\sigma}_{\mu_{n}},
\end{equation}
where $0\leq \mu_{\xi}\leq3$ for all $1\leq \xi \leq n$ are the digits of the index $\mu$ in
base-4. For example, if $\mu=33$ for a 4-qubit system,
$\hat{\Gamma}_{33}=\hat{\sigma}_{0}\otimes\hat{\sigma}_{2}\otimes\hat{\sigma}_{0}\otimes\hat{\sigma}_{1}$, since $33$ is equal to $0201$ in base-4.
For convenience, we have included a normalization constant, so that $Tr\{\hat{\Gamma}_{\mu}\hat{\Gamma}_{\nu}\}=\delta_{\mu,\nu}$ (in keeping with the
convention used in Ref.~\cite{DFVJ Measurement of qubits}).
Similarly, we write the projection operators for our measurement states as
\begin{equation}
\hat{\Pi}_{\nu}=\hat{\Pi}_{\nu_{1}}\otimes\hat{\Pi}_{\nu_{2}}\otimes...\otimes\hat{\Pi}_{\nu_{n}}.
\end{equation}
The Cholesky decomposition of $\hat{\rho}$ remains the
same, except that $T(\vec{t})$ is a $2^n\times2^n$ matrix specified by $4^n$ parameters $\vec{t}=\{t_1,t_2,\dots t_{4^n}\}$, i.e.
\begin{equation}
T(\vec{t})=
\left[\begin{array}{cccc}
    t_{1} & 0 & 0 & 0\\
    t_{2^{n}+1}+i \cdot t_{2^{n}+2} & t_{2} & 0 & 0\\
    \vdots &  & \ddots & \vdots\\
    t_{4^{n}-1}+i \cdot t_{4^{n}} & \cdots & t_{2^{n+1}-4}+i \cdot t_{2^{n+1}-3} & t_{2^{n}}
\end{array}
\right].
\end{equation}

%%%%%%%%%%%%%%%%%%%%%%%%%%%%%%%%
\subsection{Computational Constraints and Memory-efficient Linear Reconstruction\label{sub:Memory-efficient-Linear-Reconstruction}}
%%%%%%%%%%%%%%%%%%%%%%%%%%%%%%%%
In order to perform computational numerical tomography in practice, we need to take the following into consideration:

\begin{enumerate}
\item Computational efficiency - what is the upper bound on the number of
floating point operations of a certain tomography algorithm.
\item Amount of memory available - what is the upper bound on the size in
computer memory of the largest data structure used by the tomography
algorithm.
\end{enumerate}
Kronecker tensor products increase the size of resultant matrices
exponentially. The goal is to obtain the density matrix which has
$2^{n}\times2^{n}$ elements. So we cannot have any other data structure
in memory which would be larger, otherwise the problem of increasing
the number of qubits becomes constrained by that particular data structure.

For example, consider the approach described in Ref.~\cite{DFVJ Measurement of qubits}, in which
a $4^{n}\times4^{n}$ complex matrix $B_{\mu,\nu}$ (the $n$-qubit generalization of
the matrix defined by Eq.~\eqref{mightymouse}) was stored in memory. The table below outlines how much memory is needed to store
a $4^{n}\times4^{n}$ complex floating point matrix, using 32 bits
to store the real or imaginary part:
\begin{table}[H]
\begin{center}
\begin{tabular}{|c|c|c|}
\hline
Qubits&
Bytes&
GigaBytes\tabularnewline
\hline
1&
$128$&
$1.28\times10^{-7}$\tabularnewline
\hline
2&
$2.05\times10^{3}$&
$2.05\times10^{-6}$\tabularnewline
\hline
3&
$3.28\times10^{4}$&
$3.28\times10^{-5}$\tabularnewline
\hline
4&
$5.24\times10^{5}$&
$5.24\times10^{-4}$\tabularnewline
\hline
5&
$8.39\times10^{6}$&
$0.01$\tabularnewline
\hline
6&
$1.34\times10^{9}$&
$0.13$\tabularnewline
\hline
7&
$2.15\times10^{9}$&
$2.15$\tabularnewline
\hline
8&
$3.44\times10^{10}$&
$34.4$\tabularnewline
\hline
9&
$5.50\times10^{11}$&
$580$\tabularnewline
\hline
10&
$8.80\times10^{12}$&
$8.80\times10^{3}$\tabularnewline
\hline
11&
$1.41\times10^{14}$&
$1.41\times10^{5}$\tabularnewline
\hline
\end{tabular}
\end{center}

\caption{Amount of memory required to store a $4^{n}\times4^{n}$ complex
floating point matrix using 32 bits to store the real or imaginary
part. }
\end{table}

It must also be noted that any type of storage media has to be able
to perform read and write operations quite fast because this data
structure would be accessed quite frequently.
This is simply not the
case for most conventional hard drives: Using standard personal computers
of the type typically integrated into quantum optics laboratories, one is in
practice limited to about 7 qubits, without resorting to more powerful computer
hardware.  {\em However, a data structure of maximum size of $2^{n}\times2^{n}$ would
allow to go as high as 15-16 qubits, at which point the density matrix
itself would become a storage problem.}
Thus our goal is to avoid storing matrix $B_{\mu,\nu}$
into memory. Instead, we can obtain its inverse element by element.  This can be
achieved as follows: The matrix $B_{\mu,\nu}$ for an $n$-qubit system is
defined by the equation
\begin{eqnarray}
B_{\nu,\mu}&=&Tr\{{(\hat{\Pi}}_{\nu_{1}}\otimes...\otimes\hat{\Pi}_{\nu_{n}})(\hat{\sigma}_{\mu_{1}}\otimes...\otimes\hat{\sigma}_{\mu_{n}})\} \nonumber\\
&=&Tr\{\hat{\Pi}_{\nu_{1}}\hat{\sigma}_{\mu_{1}}\} Tr\{\hat{\Pi}_{\nu_{2}}\hat{\sigma}_{\mu_{2}}\}...Tr\{\hat{\Pi}_{\nu_{n}}\hat{\sigma}_{\mu_{n}}\}.
\label{eq:B-decompose}
\end{eqnarray}
Defining the $4\times 4$ matrix $\beta_{\nu_{\xi},\mu_{\xi}}=Tr\{\hat{\Pi}_{\nu_{\xi}}\hat{\sigma}_{\mu_{\xi}}\}$ for all $1\leq\xi\leq n$,  which can be easily inverted (provided a suitable set of
measurements $\{ \hat{\Pi}_{\mu}\}, \,\,(\mu=0,1,2,3)$ has been chosen),
we find
\begin{equation}
B_{\nu,\mu}^{-1}=
\beta^{-1}_{\nu_{1},\mu_{1}} \beta^{-1}_{\nu_{2},\mu_{2}}...\beta^{-1}_{\nu_{n},\mu_{n}},
\end{equation}
where as before, $(\nu_{1}, \nu_{2}, ...\nu_{n})$ are the base-4 digits of the index $\nu$ (and similarly for $\mu$).

This allows us to calculate the initial linear reconstruction of the density matrix from the observation data
$(n_0,n_1,....n_{4^n-1})$, viz:
\begin{equation}
\hat{\rho}_\mathrm{linear}=\sum_{\nu=0}^{4^{n}-1}\hat{\Gamma}_{\nu}r_{\nu},
\end{equation}
where
\begin{equation}
r_{\nu}=({\cal N})^{-1}\sum_{\mu=1}^{4^{n}-1}(B^{-1})_{\nu,\mu}n_{\mu},
\end{equation}
in a computationally efficient manner.
Now the only size constraint on linear reconstruction is the density matrix
itself. Of course, storing the projection measurement matrices $\hat{\Pi}_{\nu}$
is also problematic - a quick solution is to generate these matrices when
they become needed - one can store certain tensor combinations which
make up $\hat{\Pi}_{\nu}$ into memory and only tensor on additional combinations
to obtain the desired $\hat{\Pi}_{\nu}$.

%%%%%%%%%%%%%%%%%%%%%%%%%%%%%%%%
\subsection{Maximum Likelihood}
%%%%%%%%%%%%%%%%%%%%%%%%%%%%%%%%

Extending the Maximum Likelihood Function (MLF) from Eq.~\eqref{eq:L-fn} to $n$ qubits
we obtain:
\begin{equation}
{\cal L}(\vec{t})
=\frac{1}{2}\sum_{\nu=0}^{4^{n}-1}\frac{[{\cal N}
Tr\{\hat{\Pi}_{\nu}\hat{\rho}_\mathrm{ideal}(\vec{t})\}-n_{\nu}]^{2}}{n_{\nu}},
\end{equation}
where to simplify calculations we assumed that we can approximate variance by the measurement outcome average
in the denominator.  Minimizing this function becomes a severe computational problem. Most
gradient-free optimization routines are rather slow and only work
well for a low number of dimensions, whereas here we have a
number of dimensions which grows exponentially with the number of qubits. We have to use a numerical routine which is
more efficient; this usually involves calculating the gradient and/or
the Jacobian.  The finite-differencing approach is too slow for computing the
gradient (a fact which we verified computationally) because evaluating the MLE function
is exponentially inefficient. Hence, we require an analytic closed
form for the gradient and/or the Jacobian matrix.

It should also be noted that if the region of optimization is convex,
we are looking at non-linear convex optimization problem, for
which a number of algorithmic approaches should work. We decided to
take the simplest approach possible: Optimize the MLE function with
built-in constraints using an algorithm which works on both convex
and non-convex sets. We reduce the computation time by deriving an
analytic form for the gradient.  An alternative approach is to derive a different MLE function with
an external set of constraints, and launch another convex optimization
algorithm similar to linear programming~\cite{Kosut Convex Fidelity, Kosut Convex arXiv}.

%%%%%%%%%%%%%%%%%%%%%%%%%%%%%%%%
\subsubsection{Initial Algorithmic Attempts \label{sub:Initial-Algorithmic-Attempts}}
%%%%%%%%%%%%%%%%%%%%%%%%%%%%%%%%

The following algorithms were considered to optimize ${\cal L}$ ~\cite{Numerical Recipes,Scientific Computing},
mostly because they
are available in libraries such as GNU Scientific Library (GSL)~\cite{GSL}:

%L-fn

\begin{enumerate}
\item {} simplex method;\label{enu:simplex-Method}
\item Powell's quadratically convergent method;\label{enu:Powell's-Quadratically-Convergent}
\item Levenberg-Marquardt nonlinear least squares;\label{enu:Levenberg-Marquardt-Nonlinear-Least}
\item conjugate-gradient method;\label{enu:Conjugate-Gradient-Method}
\item BFGS algorithm.\label{enu:BFGS}
\end{enumerate}
Routines~\ref{enu:Conjugate-Gradient-Method} and~\ref{enu:BFGS} need to be able to perform gradient computation and
line search in an efficient manner and have to converge to the desired
minimum, even if started far away from it.

Let us start with the line search routines - the following algorithms
can be implemented for line search:

\begin{enumerate}
\item successive parabolic interpolation;\label{enu:Successive-Parabolic-Interpolation}
\item Newton's method;\label{enu:Newton's-Method}
\item Golden Section Search (GSS).\label{enu:Golden-Section-Search}
\end{enumerate}
In order to pick one algorithm out of these three, we need to first
know if the region of optimization is convex or not, and if so then
how closely does it resemble a quadratic function.

%%%%%%%%%%%%%%%%%%%%%%%%%%%%%%%%
\subsubsection{Matrix Calculus Derivation \label{sub:Matrix-Calculus-Derivation}}
%%%%%%%%%%%%%%%%%%%%%%%%%%%%%%%%

Regardless of the method chosen for the line search in Sec.~\ref{sub:Initial-Algorithmic-Attempts},
we still need an efficient way of computing the gradient. As established
earlier, finite-differencing requires too many function evaluations
and does not satisfy our computational efficiency constraints.

The goal of this section is to find the gradient of ${\cal L}$ in
closed form. This procedure can then be extended to finding second
order partial derivatives for the Hessian matrix and differentiation
with respect to a constant for line search routine.

We begin with the gradient derivation. This reduces to finding
\begin{equation}
\frac{\partial{\cal L}}{\partial t_{\nu}}=\frac{\partial{\cal L}}{\partial T}\frac{\partial T}{\partial t_{\nu}}.
\end{equation}
Using matrix calculus, it suffices to find $\frac{\partial{\cal L}}{\partial T}$,
which would be a matrix of size $2^{n}\times2^{n}$ in our case. Certain
elements of this matrix would represent the values of $\frac{\partial{\cal L}}{\partial t_{\nu}}$:

\begin{eqnarray}
&&\frac{\partial{\cal L}(\vec{t})}{\partial T} =
\frac{\cal N}{Tr\{ T^{\dagger}(\vec{t})T(\vec{t})\}^{2}}
\sum_{\nu=0}^{4^{n}-1}\left[\frac{{\cal N}Tr\{\hat{\Pi}_{\nu}\hat{\rho}_\mathrm{ideal}(\vec{t})\}-n_{\nu}}{n_{\nu}}\right]\nonumber\\
&&\hspace{5mm}\times\left[
Tr\{ T^{\dagger}(\vec{t})T(\vec{t})\}
\frac{\partial Tr\{\hat{\Pi}_{\nu}T^{\dagger}(\vec{t})T(\vec{t})\}}{\partial T}\right.\nonumber\\
&&\left.
\hspace{20mm}-Tr\{\hat{\Pi}_{\nu}T^{\dagger}(\vec{t})T(\vec{t})\}
\frac{\partial Tr\{ T^{\dagger}(\vec{t})T(\vec{t})\}}{\partial T}
\right]
\label{eq:dAlpha by dT}
\end{eqnarray}

Defining the following real quantities:
\begin{eqnarray}
A(\vec{t})&=&Tr\{ T^{\dagger}(\vec{t})T(\vec{t})\}~\mathrm{and}\\
B_{\nu}(\vec{t})&=&Tr\{\hat{\Pi}_{\nu}T^{\dagger}(\vec{t})T(\vec{t})\},
\end{eqnarray}
we find:
\begin{equation}
Tr\{\hat{\Pi}_{\nu}\hat{\rho}_\mathrm{ideal}(\vec{t})\}=\frac{B_{\nu}}{A}.
\end{equation}
Further we will denote the matrix derivatives of these quantities with respect to the Cholesky matrix $T$ as follows:
\begin{eqnarray}
B_{\nu}^{\prime}(\vec{t})&=&
\displaystyle{\frac{\partial Tr\{\hat{\Pi}_{\nu}T^{\dagger}(\vec{t})T(\vec{t})\}}{\partial T}},\\
&&\nonumber\\
A^{\prime}(\vec{T})&=&
\displaystyle{\frac{\partial Tr\{ T^{\dagger}(\vec{t})T(\vec{t})\}}{\partial T}}.
\end{eqnarray}
Because matrix calculus is only well-defined for real-valued matrices, let us write
\begin{equation}
T=T(\vec{t})=X+i\cdot Y,\qquad\hat{\Pi}_{\nu}=K_{\nu}+i\cdot\Lambda_{\nu}.\end{equation}
Then, using the matrix calculus theorems in Sec.~\ref{sec:Matrix-Differential-Calculus}
we find
\begin{eqnarray}
A^{\prime}&=&2X+i\cdot2Y,\\
B_{\nu}^{\prime}&=&2XK_{\nu}-2Y\Lambda_{\nu}+i\cdot(2X\Lambda_{\nu}+2YK_{\nu}).
\end{eqnarray}
Denoting
\begin{equation}
C_{\nu}=\left[\frac{NB_{\nu}-An_{\nu}}{An_{\nu}}\right],\qquad D_{\nu}=\left[\frac{AB_{\nu}^{\prime}-B_{\nu}A^{\prime}}{A^{2}}\right]
\end{equation}
we find the matrix derivative of ${\cal L}$ can be written in the compact form
\begin{equation}
\frac{\partial{\cal L}}{\partial T}={\cal L}^{\prime}(\vec{t})={\cal N}\sum_{\nu=0}^{4^{n}-1}C_{\nu}D_{\nu},\label{eq:alpha first derivative}
\end{equation}
where $C_{\nu}$ is a scalar and $D_{\nu}$ is a $2^{n}\times2^{n}$
matrix.
In fact, the upper-diagonal of ${\cal L}'(\vec{t})$ and imaginary
part of the diagonal are of no use to us - values of the gradient
are seeded in the original locations of $t_{\nu}$, so ${\cal L}'(\vec{t})$
has to be disassembled into real and imaginary parts and then the
gradient vector has to be filled from the resulting matrices.

%%%%%%%%%%%%%%%%%%%%%%%%%%%%%%%%%%%%%%%%%%%%%%%%%%%%%%%%%%%%%%%%
%%%%%%%%%%%%%%%%%%%%%%%%%%%%%%%%%%%%%%%%%%%%%%%%%%%%%%%%%%%%%%%%
%%%%%%%%%%%%%%%%%%%%%%%%%%%%%%%%%%%%%%%%%%%%%%%%%%%%%%%%%%%%%%%%
\section{Description of Code \label{sec:Description-of-Code}}
%%%%%%%%%%%%%%%%%%%%%%%%%%%%%%%%%%%%%%%%%%%%%%%%%%%%%%%%%%%%%%%%
%%%%%%%%%%%%%%%%%%%%%%%%%%%%%%%%%%%%%%%%%%%%%%%%%%%%%%%%%%%%%%%%
%%%%%%%%%%%%%%%%%%%%%%%%%%%%%%%%%%%%%%%%%%%%%%%%%%%%%%%%%%%%%%%%
The goal is to scale tomography routines up to a higher number of
qubits on a standard single-processor workstation by refining the
tomography algorithms to remove the numerical complexity bottleneck from experimental
post-processing. In this section, we describe how the codes were implemented.

%%%%%%%%%%%%%%%%%%%%%%%%%%%%%%%%
\subsection{State Tomography Routine}
%%%%%%%%%%%%%%%%%%%%%%%%%%%%%%%%

Four routines provide tomography and run in the following order:

\begin{enumerate}
\item Linear reconstruction \label{enu:Linear-Reconstruction} - provides
the linear reconstruction of the data by inverting the measurements
into a matrix $\hat{\rho}_\mathrm{linear}$, outlined in Sec.~\ref{sub:Memory-efficient-Linear-Reconstruction},
which has all characteristics of a density matrix, except positive
semi-definiteness.
\item ``Quick and Dirty" \label{enu:Quick-and-Dirty} - quickly fixes
$\hat{\rho}_\mathrm{linear}$ into $\hat{\rho}_\mathrm{QD}$ by setting all negative
eigenvalues to zero and re-normalizing.
\item ``Forced Purity" - for pure states, eigenvalues are known. This routine
forces eigenvalues of $\hat{\rho}_\mathrm{linear}$ or $\hat{\rho}_\mathrm{QD}$
(does not matter which one) into those of a pure state, also ensuring unit-trace condition.
\item MLE \label{enu:Conjugate-Gradient-Optimization} - we use
the elements of the ``Quick and Dirty" density matrix as a starting point
for our optimization routine. We then launch the BFGS2 algorithm supplied
with GSL.
\end{enumerate}
Our progress while developing these routines is as follows:

\begin{enumerate}
\item Started with our own simplex method code in Matlab, which only
optimized 4 qubits - gradient-based algorithm was needed.
\item Wrote the conjugate-gradient routine in Matlab using GSS line search routine
and using finite-difference gradient, which allowed for 5-6 qubit
tomography.
\item Applied matrix differential calculus to the gradient and obtained
a closed form expression, which severely improved Matlab routines
for up to 7 qubits.
\item Experimented with Newton's method and successive parabolic interpolation
line searches which did not work in the end. This led us to suspect
that the region of optimization is not convex.
\item Re-wrote everything in C using GSL and employed GSL's BFGS2 algorithm
and its collection of line searches - this pushed our routines to
9 qubits (MLE limits to 9 qubits, but not ``Forced Purity").
\end{enumerate}
All code is currently implemented in C using GSL, with prototype routines
also available in Matlab.

%%%%%%%%%%%%%%%%%%%%%%%%%%%%%%%%
\subsection{Creating Pseudo-Experimental Data}
%%%%%%%%%%%%%%%%%%%%%%%%%%%%%%%%

Generally, if one wants to simulate a physical
state characterized by $\hat{\rho}_\mathrm{physical}$
with $100\epsilon\%$ experimental state error, then

\[
\hat{\rho}_\mathrm{physical}=(1-\epsilon)\cdot\hat{\rho}_\mathrm{theoretical}+\epsilon\cdot\hat{\rho}_\mathrm{random},\]
where $\hat{\rho}_\mathrm{theoretical}$ is a density matrix of some desired
state and $\epsilon$, a real-valued constant, simulates experimental {}``state error'' - the physical state always differs from the intended state by some small amount; a random density matrix is created as follows:
\begin{equation}
R=2\cdot rand(2^{n})-1+i\cdot(2\cdot rand(2^{n})-1)\label{eq:R random},\end{equation}
\begin{equation}
\hat{\rho}_\mathrm{random}=\frac{R^{\dagger}R}{Tr\{ R^{\dagger}R\}},\label{eq:rho random}\end{equation}
where \emph{rand} function creates a $2^{n}\times2^{n}$ matrix of
pseudo-random values, sampled from Uniform(0,1) distribution \footnote{See the Matlab online manual at http://www. mathworks. com .}.

For instance, the following results in a noisy GHZ state:
\[
\hat{\rho}_\mathrm{GHZ}=(1-\epsilon)\cdot\frac{1}{2}|100...01\rangle\langle100...01|+\epsilon\cdot\hat{\rho}_\mathrm{random}.\]
The simulation routine creates a physical density matrix,
simulates experimental measurement outcomes and then attempts to reconstruct
this density matrix. Knowing what the reconstructed density matrix
should be, enables us to compare how well each reconstruction routine
works for a certain number of qubits.

The expected number of positive outcomes is obtained using Eq.~\eqref{eq:n sub nu from density matrix}, viz:
\[
\bar{n}_{\nu}={\cal N}Tr\{
\hat{\Pi}_{\nu_{1}}\otimes\hat{\Pi}_{\nu_{2}}\otimes...\otimes\hat{\Pi}_{\nu_{n}}
\hat{\rho}_\mathrm{physical}\},
\]
where ${\cal N}$ is a constant which is equivalent to the number of times repeated projective measurements were taken
\footnote{${\cal N}$ was set to $104$ in our tomographic routine, but it can
be any positive integer as long as $n_{\nu}$ values resemble realistic photon
counts, and not fractions less than 1. %
}. We then add experimental noise to the measurements %
\footnote{Note that if the measurements are performed with zero noise, then
the Linear Reconstruction routine performs tomography of the density
matrix with Fidelity value of 1.%
} using:
\[
n_{\nu}=Poisson(\bar{n}_{\nu}),\]
where $Poisson(\lambda)$ generates a random number from a Poisson
distribution with mean $\lambda$ using a probability integral transformation.

%%%%%%%%%%%%%%%%%%%%%%%%%%%%%%%%%%%%%%%%%%%%%%%%%%%%%%%%%%%%%%%%
%%%%%%%%%%%%%%%%%%%%%%%%%%%%%%%%%%%%%%%%%%%%%%%%%%%%%%%%%%%%%%%%
%%%%%%%%%%%%%%%%%%%%%%%%%%%%%%%%%%%%%%%%%%%%%%%%%%%%%%%%%%%%%%%%
\section{Results \label{sec:Results}}
%%%%%%%%%%%%%%%%%%%%%%%%%%%%%%%%%%%%%%%%%%%%%%%%%%%%%%%%%%%%%%%%
%%%%%%%%%%%%%%%%%%%%%%%%%%%%%%%%%%%%%%%%%%%%%%%%%%%%%%%%%%%%%%%%
%%%%%%%%%%%%%%%%%%%%%%%%%%%%%%%%%%%%%%%%%%%%%%%%%%%%%%%%%%%%%%%%

In this section, we discuss the conclusions to be drawn from the numerical trials
described in  the previous sections.  In particular we address the question posed in the title of this paper: Do we
always need an expensive MLE routine to perform tomography or would
{}``Quick and Dirty'' or {}``Forced Purity'' methods suffice?

We compare the {}``Quick and Dirty'' and {}``Forced Purity'' routines
to the {}``MLE'' routine for states with wide variations of entropy and
entanglement. We also show how well these routines scale in runtime and
how experimental errors affect the reconstructed states as the number
of qubits increases.

The linear entropy, which specifies the degree of purity of the state, is defined as
\[
S_\mathrm{linear}(\hat{\rho})=\frac{2^{n}}{2^{n}-1}[1-Tr\{\hat{\rho}\}]\]
for n qubits.

The tangle (i.e. the square of the concurrence~\cite{Wootters}) is  defined for 2 qubits,
as
\[
\tau=[max\{\lambda_{4}-\lambda_{1}-\lambda_{2}-\lambda_{3},0\}]^{2},\]
where $\lambda$'s are the square roots of the eigenvalues of the
matrix $\sqrt{\hat{\rho}}(\hat{\sigma}_{y}\otimes\hat{\sigma}_{y})\hat{\rho}^{*}(\hat{\sigma}_{y}\otimes\hat{\sigma}_{y})\sqrt{\hat{\rho}}$,
which is guaranteed to be Hermitian~\cite{Wootters} and $\hat{\sigma}_{y}\otimes\hat{\sigma}_{y}$
is the spin-flip matrix and $\hat{\rho}^{*}$ is the complex conjugate
of density matrix $\hat{\rho}$.  For larger numbers of qubits, it can be
used as a lower bound on the degree of entanglement~\cite{key-26}.

%%%%%%%%%%%%%%%%%%%%%%%%%%%%%%%%
\subsection{Linear Entropy vs Tangle Plane\label{sub:Linear-Entropy-vs}}
%%%%%%%%%%%%%%%%%%%%%%%%%%%%%%%%

We observed that for 2 qubits, certain states produce fidelities of
over 90\% using ``Quick and Dirty" routine and consistently high fidelities
using MLE. We generated $2\times10^{6}$ pseudo-random density matrices,
that filled the entire entropy-tangle plane. Random density matrices
had to be biased in order to evenly cover the entire plane. For example,
to fill the plane below the Werner state line, we used:

\begin{equation}
\hat{\rho}_\mathrm{\tau}=\left[\begin{array}{cccc}
1-\delta^{2} & 0 & 0 & \delta\sqrt{1-\delta^{2}}\\
0 & 0 & 0 & 0\\
0 & 0 & 0 & 0\\
\delta\sqrt{1-\delta^{2}} & 0 & 0 & \delta^{2}\end{array}\right],\label{eq:pure_state}\end{equation}
which biased the tangle in

\[
\hat{\rho}_\mathrm{trial}=\epsilon^{2}\cdot\hat{\rho}_\mathrm{random}+(1-\epsilon^{2})\cdot\hat{\rho}_\mathrm{\tau},\]
where

\[
0\leq\epsilon\leq1,\qquad0\leq\delta\leq1/\sqrt{2}.\]

Varying $\delta$ changes tangle and varying $\epsilon$ changes entropy.
We cycled through $100\times100$ different combinations of $\delta$
and $\epsilon$ and for each setting performed 100 trials, sampling
$\hat{\rho}_\mathrm{random}$ from Uniform(-1,1) probability distribution
for each trial. We can also move along the Maximally Entangled Mixed
State (MEMS) line by varying $\gamma$,
\[
\hat{\rho}_\mathrm{MEMS}=\left[\begin{array}{cccc}
g(\gamma) & 0 & 0 & \gamma/2\\
0 & 1-2g(\gamma) & 0 & 0\\
0 & 0 & 0 & 0\\
\gamma/2 & 0 & 0 & g(\gamma)\end{array}\right],\]
where
\[
g(\gamma)=\left\{ \begin{array}{c}
\gamma/2,\;\gamma\geq2/3\\
1/3,\;\gamma<2/3\end{array}\right.\]
and
\[
\hat{\rho}_\mathrm{trial}=\epsilon^{2}\cdot\hat{\rho}_\mathrm{random}+(1-\epsilon^{2})\cdot\hat{\rho}_\mathrm{MEMS}.\]
We further sampled $1000\times1000$ different settings of $\epsilon$
and $\gamma$ to fill the area around MEMS line: In this case increasing
$\epsilon$ increases the distance from the MEMS line.

\begin{figure}[H]
\includegraphics[width=0.9 \columnwidth]{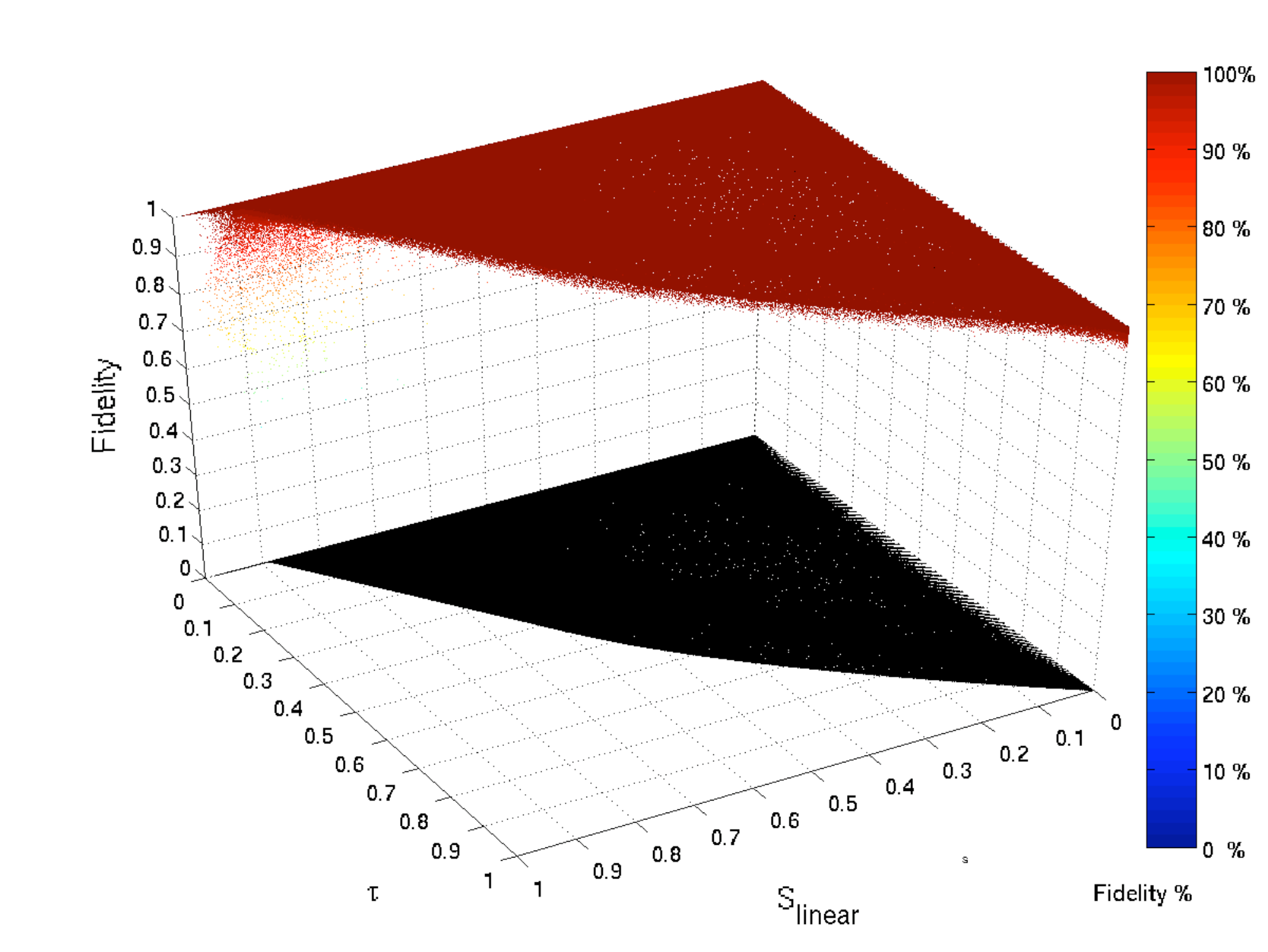}
\caption{On the $S_\mathrm{linear}$ and $\tau$ plane you can see that the $2\times10^{6}$
generated states covered the entire plane; above each point is the
corresponding fidelity of the recovered state using maximum likelihood.
Notice that most fidelity values lie between 90\%-99\%. The fidelity
values create a thin plane, which suggests that the standard deviation
is low for various states. However, some high-entropy states cannot
be recovered well even with the expensive MLE procedure.}
\end{figure}
\begin{figure}[H]
\includegraphics[width=0.9 \columnwidth]{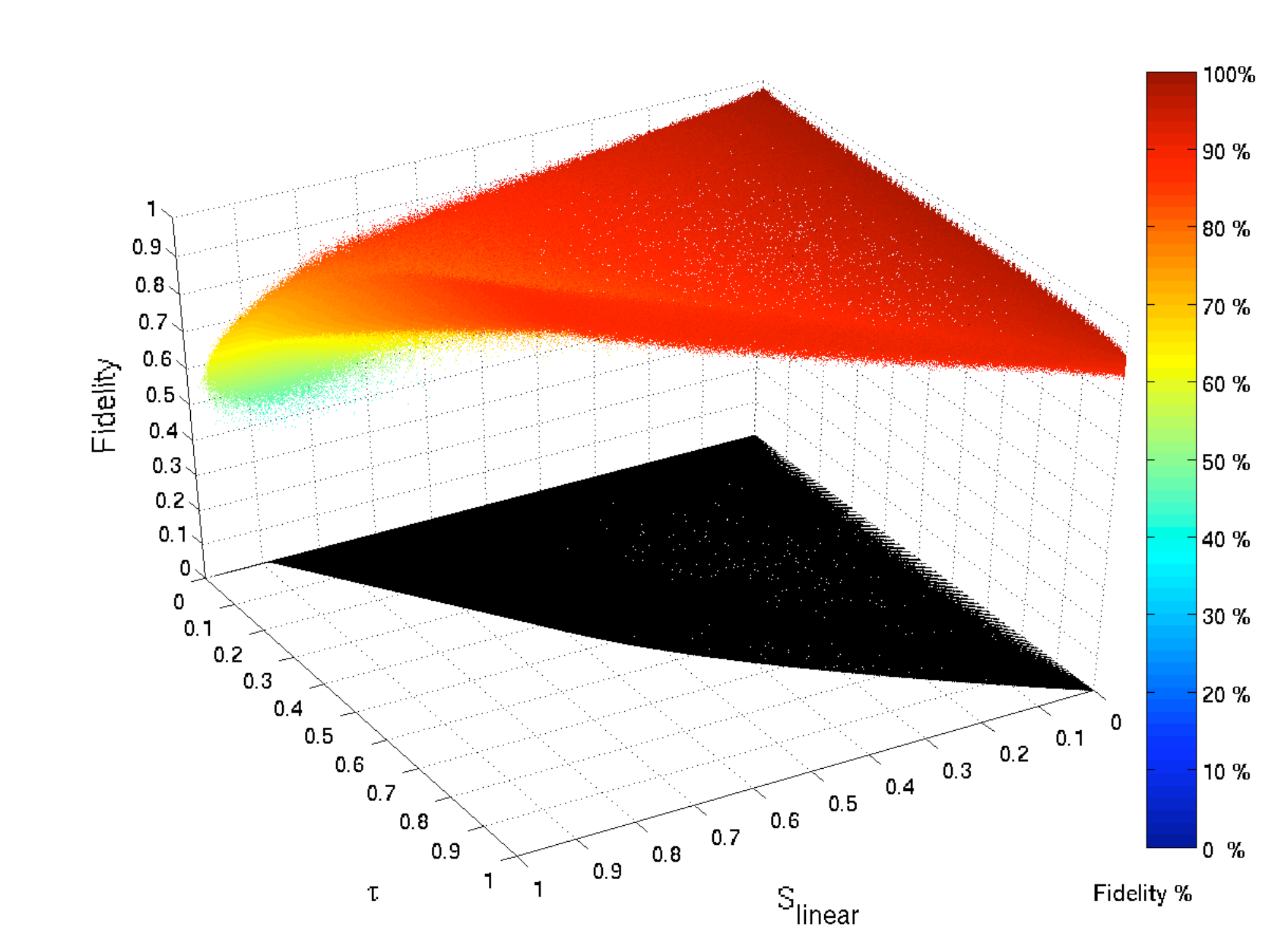}
\caption{These are the same $2\times10^{6}$ states as in the previous figure
and the projection on $S_\mathrm{linear}$ and $\tau$ plane is identical.
Notice how the ``Quick and Dirty" routine also fails to reach high fidelity
values as the states become more mixed. However, for pure states Quick
and Dirty is comparable to MLE in fidelity values.}
\end{figure}

This suggests that for states with low entropies (pure states), Quick
and Dirty routine should work in theory. This is not surprising, as
from the spectral decomposition, we can see that all states with $S_\mathrm{linear}=0$
property, regardless of the value of $\tau$, share one thing in common:
eigenvalues. More precisely, for n qubits, eigenvalue 0 occurs with
degeneracy $2^{n}-1$ and eigenvalue 1 occurs with degeneracy $1$.
So, setting negative eigenvalues to zero adjusts the eigenvalues closer
to the eigenvalues of a pure state. If we know that the state is pure
ahead of time, we can just reset the eigenvalues to the known values
after the linear inversion procedure and obtain the density matrix
- this is further explored in Sec.~\ref{sub:Forced-Tomography-for}.
\begin{figure}[H]
\includegraphics[width=0.9 \columnwidth]{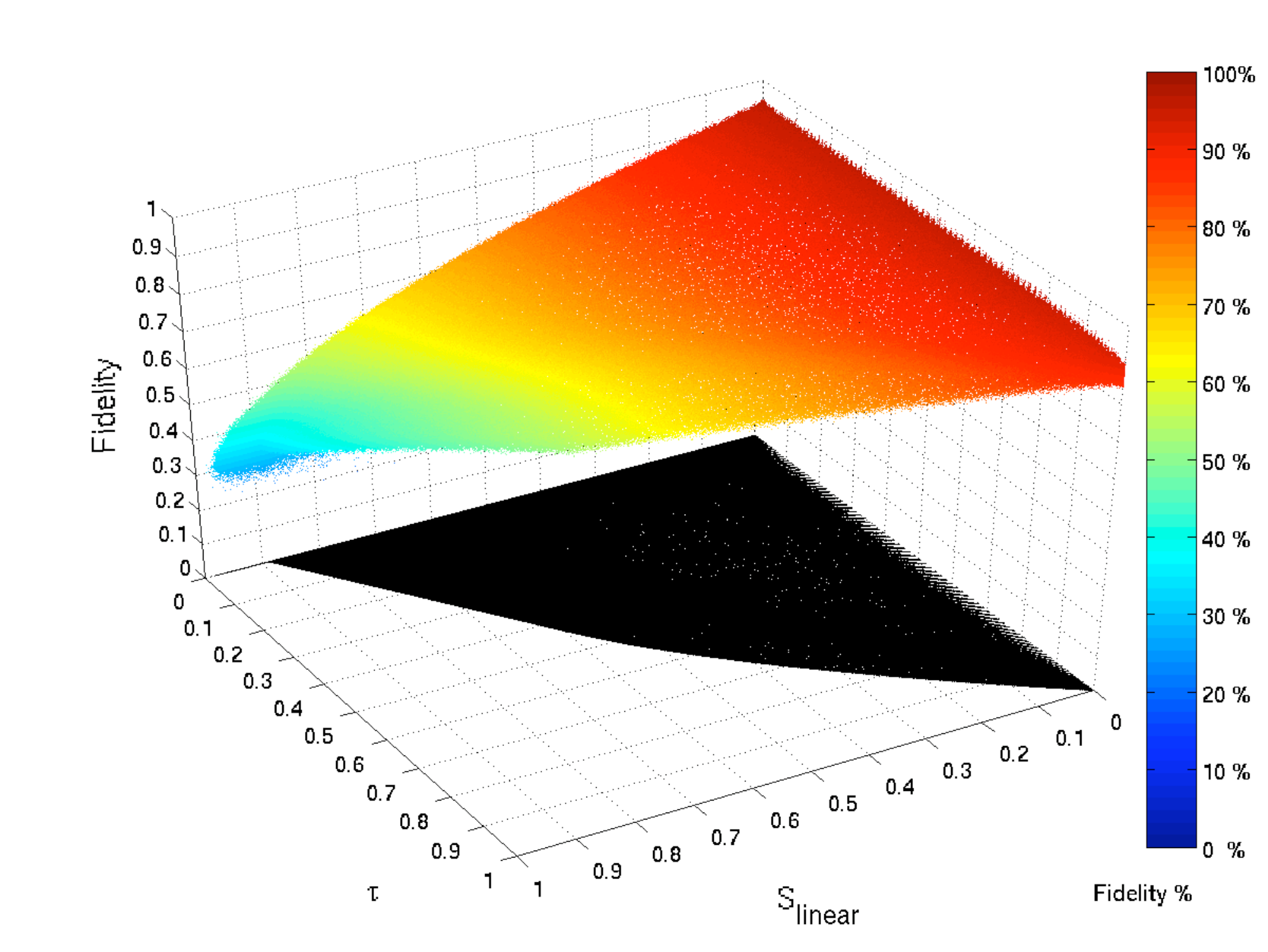}
\caption{Again, same $2\times10^{6}$ states, but this time with ``Forced Purity"
performed. Notice that for 2 qubits ``Forced Purity" appears to be worse
than ``Quick and Dirty" even for pure states - this is not the case when
the number of qubits increases, as we explore in the next section.}
\end{figure}

%%%%%%%%%%%%%%%%%%%%%%%%%%%%%%%%
\subsection{Performance for N qubits }
%%%%%%%%%%%%%%%%%%%%%%%%%%%%%%%%
In order to extend this assessment to larger numbers of qubits,
while still varying the amount of entanglement and disorder,
we considered a generalized version of the Werner state for $n$ qubits. Since this state slices
through the entire plane presented in Sec.~\ref{sub:Linear-Entropy-vs},
we can see how well tomography operates on states with various tangle
and entropy values by varying the location along the Werner state
line. An adjusted Werner state density matrix is given by:
\[
\hat{\rho}_\mathrm{Werner}=|\mathrm{GHZ}\rangle\langle \mathrm{GHZ}|\cdot\epsilon+\frac{(1-\epsilon)}{2^n} \cdot \hat{I},\]
where
\[
|\mathrm{GHZ}\rangle=\frac{1}{\sqrt{2}}\{|00...0\rangle+|11...1\rangle\}\]
and $\hat{I}$ is the $2^{n}\times2^{n}$ identity operator, representing
a maximally mixed state. When $\epsilon=1$ we obtain a state located
at $S_\mathrm{linear}=0$ and $\tau=1$ and when $\epsilon=0$ we obtain $S_\mathrm{linear}=1$
and $\tau=0$. We then vary $\epsilon$ from $0$ to $1$ in $101$
increments and for each value of $\epsilon$ perform tomography $100$
times, for a fixed number of qubits.

\begin{figure}[H]
\includegraphics[width=0.8 \columnwidth]{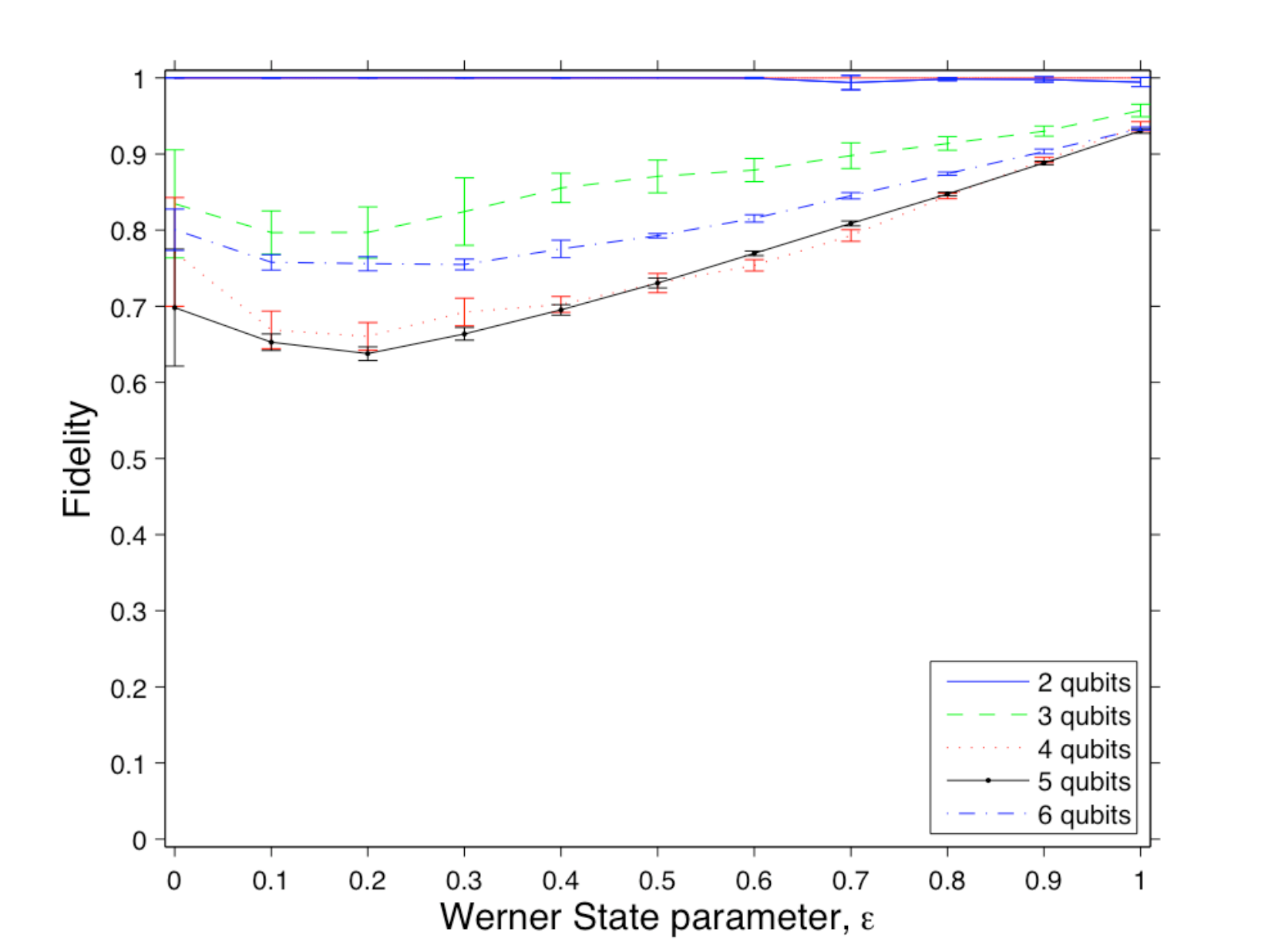}
\caption{MLE works well for 2 qubits, but for a higher number of qubits there
is a significant drop in fidelity. A 10\% state error is quite large,
but even at this error pure states are reconstructed better than mixed
states as you can see around $\epsilon=1$ (highly entangled pure
state).}
\end{figure}

\begin{figure}[H]
\includegraphics[width=0.8 \columnwidth]{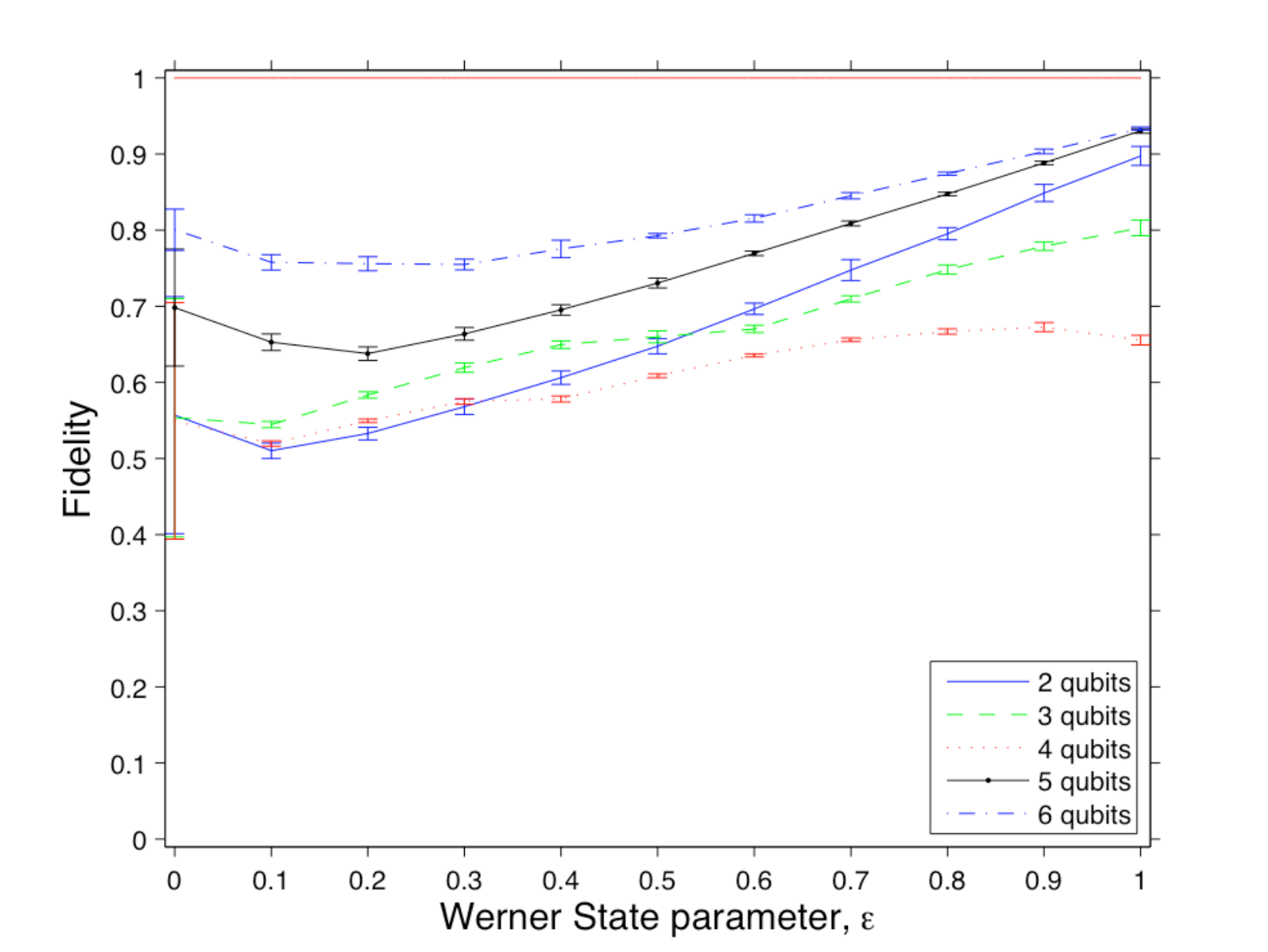}
\caption{``Quick and Dirty" routine is not performing very well even for pure
states, although for a certain number of qubits it appears to work.
On the contrary, it appears to be improving as the number of qubits
is increasing.}
\end{figure}

\begin{figure}[H]
\includegraphics[width=0.8 \columnwidth]{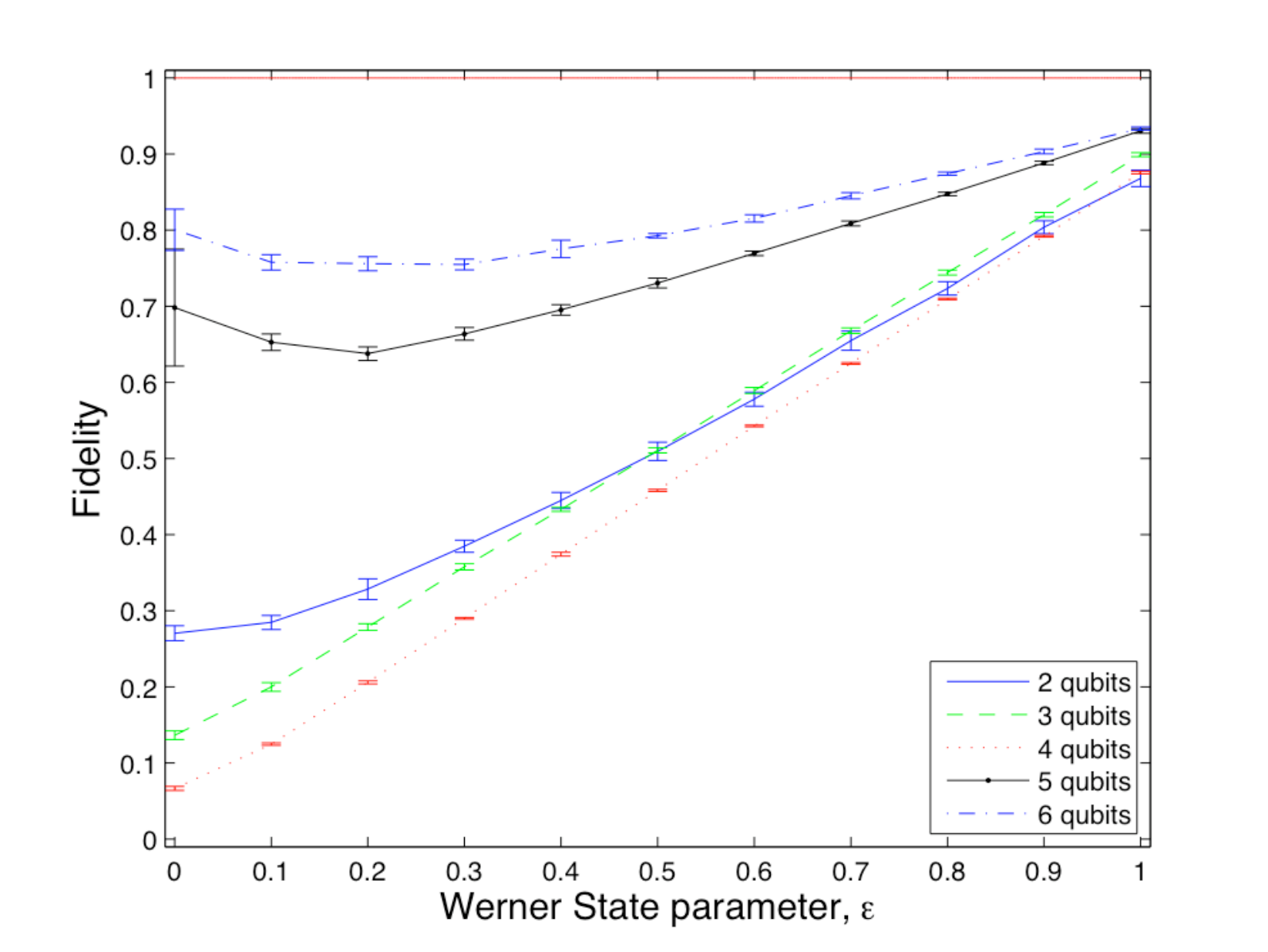}
\caption{For ``Forced Purity" we can confidently say that the routine is improving
significantly, as the number of qubits is increasing. And for pure
entangled states this routine is almost as good as MLE, but runs at
a fraction of the time!}
\end{figure}

%%%%%%%%%%%%%%%%%%%%%%%%%%%%%%%%
\subsection{``Forced Purity" Tomography for Pure States\label{sub:Forced-Tomography-for}}
%%%%%%%%%%%%%%%%%%%%%%%%%%%%%%%%%%%%%%%%%%%%%%%%%%

In order for {}``Forced Purity'' to work, the measurement
outcomes have to be sufficiently close to their true values.
To address the issue of `how close', we simulated
pure states, with 11 distinct tangle values evenly spaced between $0$ and $1$ and then started with ${\cal N}=10$
- i.e. ${\cal N}$ repeated projective measurements for each measurement outcome. We
then increased ${\cal N}$ by one at each iteration and repeated ``Forced Purity" tomography
for some number of qubits. As soon as ${\cal N}$ allowed ``Forced Purity"
to perform tomography at 90\% fidelity, the routine was terminated and
the value for ${\cal N}$ recorded.

One of the possible reasons why MLE tomography did not yield high fidelity values
for a larger number of qubits is that it also required more accurate estimates of the measurement outcomes.
Because MLE produced almost equal fidelities to ``Forced Purity"
for pure states, we would expect the same number of ${\cal N}$ to work
for MLE tomography.

%%%%%%%%%%%%%%%%%%%%%%%%%%%%%%%%
\subsection{Runtime Analysis}
%%%%%%%%%%%%%%%%%%%%%%%%%%%%%%%%
Section~\ref{sub:Forced-Tomography-for} suggests that pure states
do not require expensive MLE techniques for tomography. Nonetheless,
it is interesting to see how MLE scales in runtime compared to Quick
and Dirty and ``Forced Purity" routines. Here we present runtimes and
fidelity estimates for a pure state with $\tau\approx0.5$ and a slightly
mixed state with the same tangle value which lies on the Werner state
line. We also show that even the expensive MLE routine decreases in
fidelity as we increase the number of qubits. For this analysis we
assume that an experiment can be performed a sufficiently large number
of times, $10^{6}$ to be exact.

In conclusion, {}``Forced Purity'' results in lower Fidelity values for 2 qubits
than {}``Quick and Dirty'', but then increases in Fidelity and converges
to MLE's fidelity for a higher number of qubits.

\begin{table}[H]
\begin{center}
\begin{tabular}{|c|c|c|c|c|}
\hline
n&
MLE&
Iteration Time&
Q \& D&
FP\tabularnewline
\hline
2&
$40\pm10$&
$1.0\pm0.5$&
$0.19\pm0.05$&
$0.17\pm0.04$\tabularnewline
\hline
3&
$1000\pm300$&
$12\pm8$&
$0.29\pm0.03$&
$0.26\pm0.02$\tabularnewline
\hline
4&
$(16\pm3)\,103$&
$180\pm70$&
$1.1\pm0.1$&
$1.1\pm0.2$\tabularnewline
\hline
5&
$(35\pm6)\,104$&
$(4\pm1)\,103$&
$9.4\pm4$&
$12\pm4$\tabularnewline
\hline
6&
$(6.0\pm0.3)\,106$&
$(60\pm16)\,103$&
$58\pm2.5$&
$60\pm5.4$\tabularnewline
\hline
\end{tabular}
\end{center}

\caption{Runtimes in milliseconds for a state at $\tau\approx0.5$ and $S_\mathrm{linear}=0$. ``Iteration Time" measures how long the line search routine takes for each iteration of the BFGS routine.
Abbreviations: MLE: Complete Maximum Likelihood reconstruction; QD: ``Quick and Dirty" method; FP: ``Forced Purity".}
\end{table}

\begin{table}[H]
\begin{center}
\begin{tabular}{|c|c|c|c|c|}
\hline
n&
MLE&
Iteration Time&
Q \& D&
FP\tabularnewline
\hline
2&
$41\pm10$&
$0.77\pm0.4$&
$0.060\pm0.005$&
$0.038\pm0.006$\tabularnewline
\hline
3&
$1200\pm200$&
$12\pm20$&
$0.17\pm0.003$&
$0.36\pm0.6$\tabularnewline
\hline
4&
$(19\pm1)\, 103$&
$200\pm100$&
$1.3\pm0.4$&
$1.1\pm0.1$\tabularnewline
\hline
5&
$(29\pm2)\,104$&
$2900\pm900$&
$7.6\pm0.6$&
$8.8\pm2$\tabularnewline
\hline
6&
$(6.7\pm0.4)\,106$&
$(7\pm2)\,104$&
$65\pm5$&
$67\pm5$\tabularnewline
\hline
\end{tabular}
\end{center}
\caption{Runtimes in milliseconds for a state at $\tau\approx0.5$ and $S_\mathrm{linear}$
to that along the Werner state line; abbreviations same as Table II. }
\end{table}

\begin{figure}[H]
\includegraphics[width=0.8 \columnwidth]{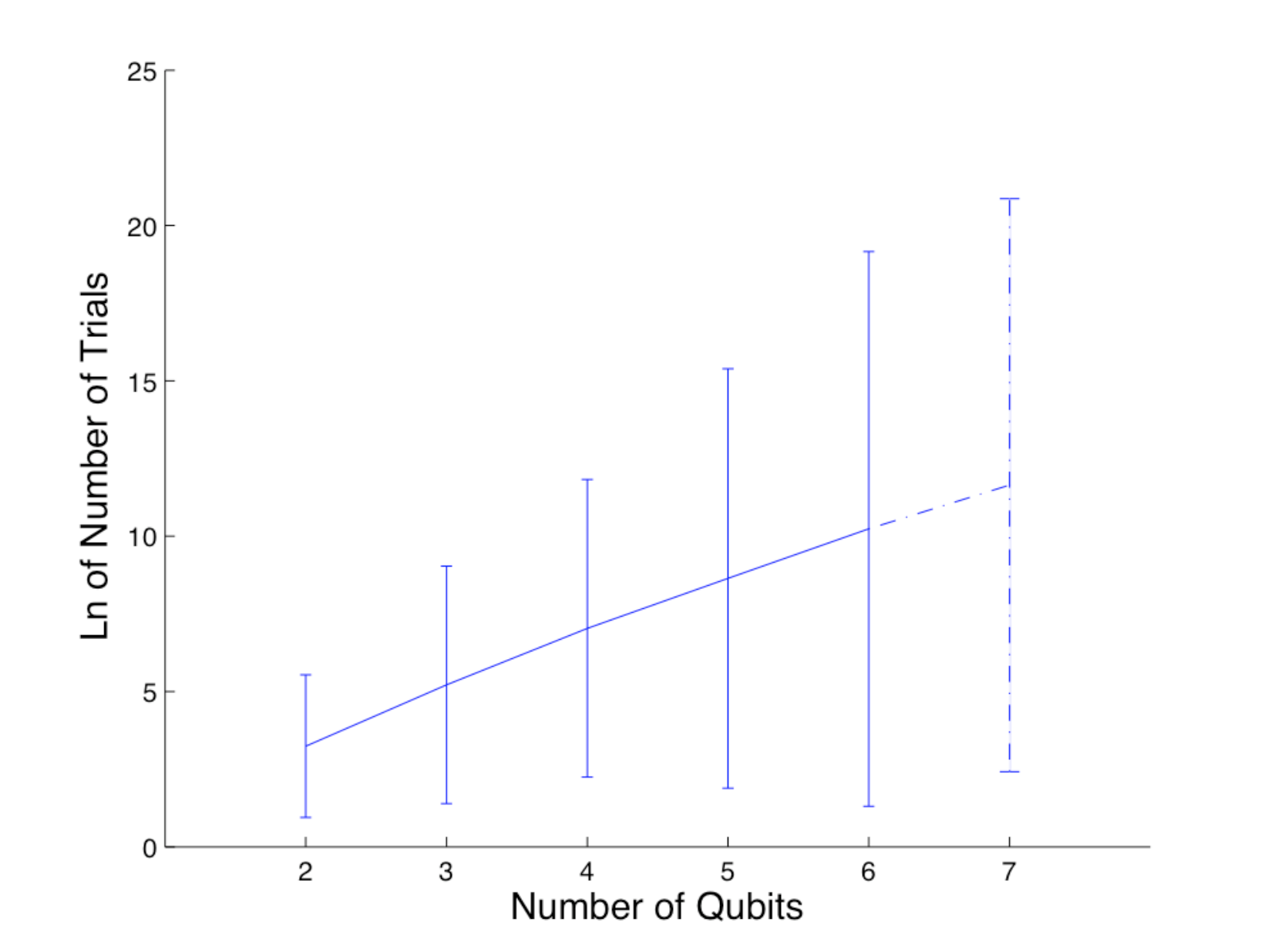}
\caption{For each number of qubits, this plot shows the number of times each
projection measurement has to be repeated for a state, in order to obtain an accurate estimate of the
measurement outcome. With a 5\% state error, this number of measurements will allow the Forced
Purity routine to estimate the state with 90\% fidelity. There is an exponential increase in how many times the experiment has to be repeated. }
\end{figure}
\begin{figure}[H]
\includegraphics[width=0.8 \columnwidth]{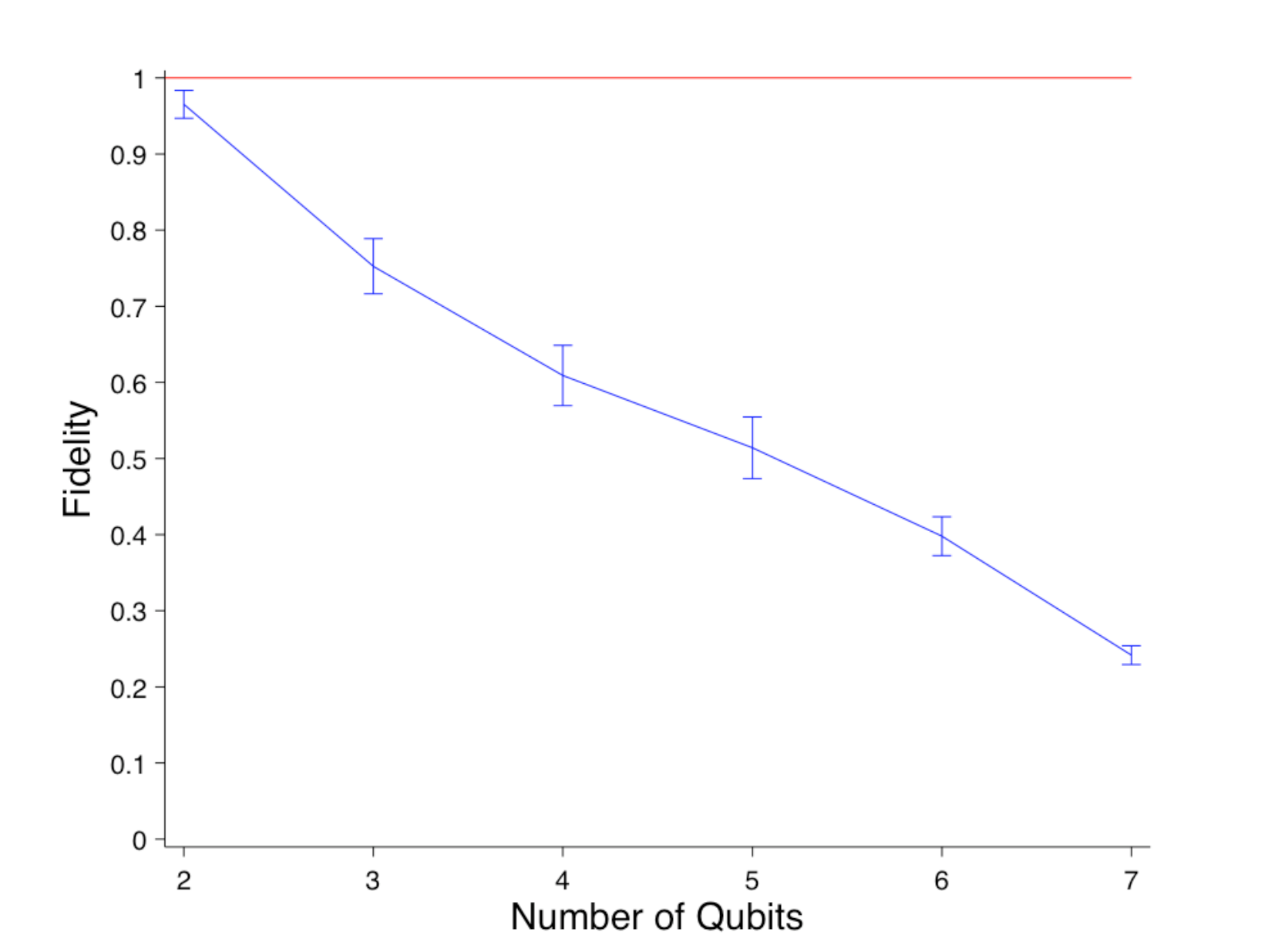}
\caption{\label{fig:QD_tau_drop}
This graph shows the average fidelities of estimated pure states recovered using
the ``Quick and Dirty" approach for different numbers of qubits at 5\% state error.  In total 11 tangle values
equally spaced between 0 and 1 were used, and 10 recoveries performed for each tangle value.
This shows that while ``Quick and Dirty" appears
to work for 2 qubits, in the long run it linearly worsens and cannot
be used as a suitable tomography algorithm.}
\end{figure}
\begin{figure}[H]
\includegraphics[width=1.0 \columnwidth]{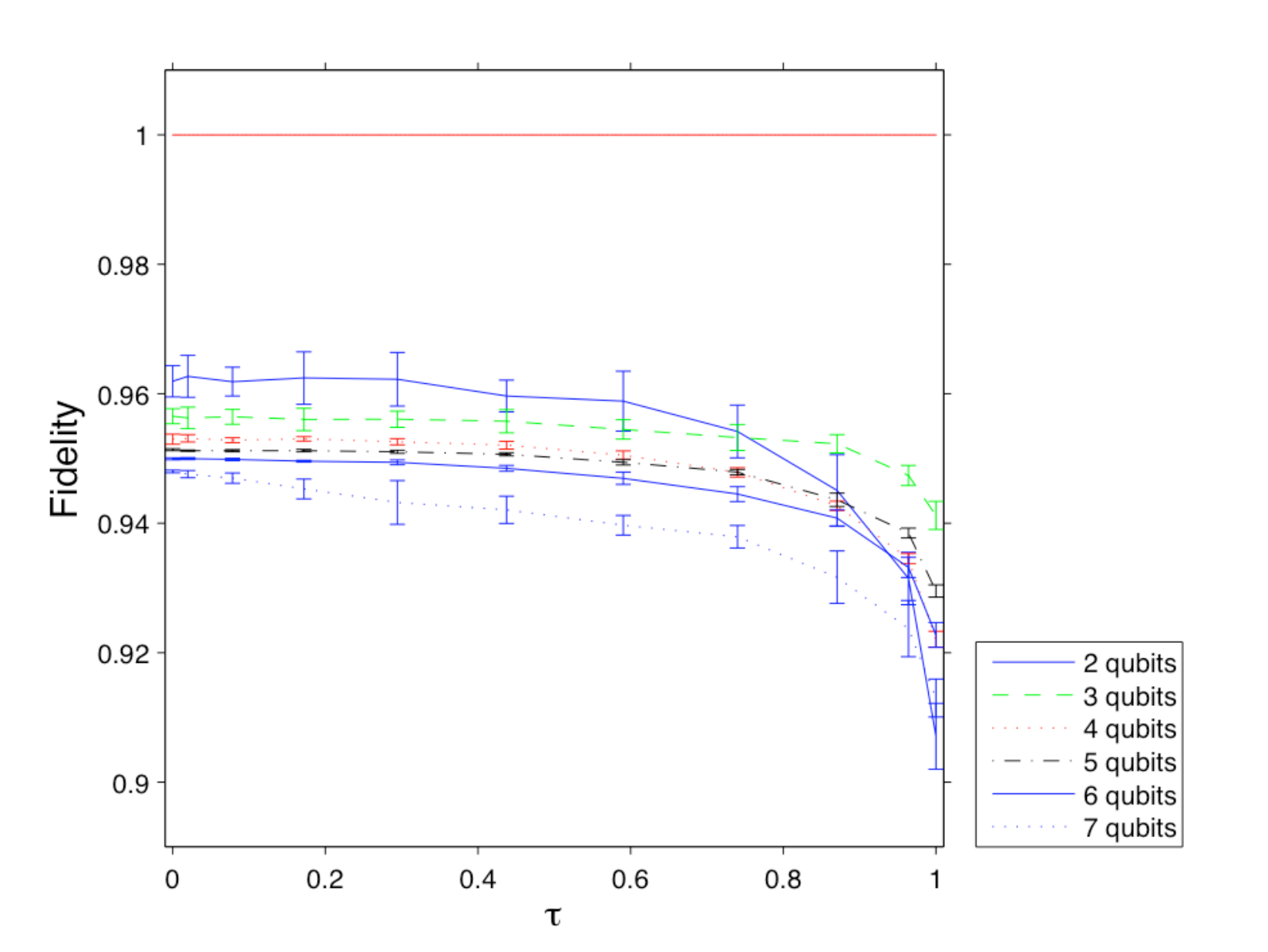}
\caption{Using the same procedure as in Fig.~\ref{fig:QD_tau_drop} we then
performed ``Forced Purity" on each state. We said that for the Werner
state line ``Forced Purity" routine improves overall as the number of
qubits increases. This is not the case for pure states, and as we
can clearly see ``Forced Purity" slightly decoheres, but overall still
remains at over 90\% fidelity even for 7 qubits.}
\end{figure}

%%%%%%%%%%%%%%%%%%%%%%%%%%%%%%%%%%%%%%%%%%%%%%%%%%%%%%%%%%%%%%%%
%%%%%%%%%%%%%%%%%%%%%%%%%%%%%%%%%%%%%%%%%%%%%%%%%%%%%%%%%%%%%%%%
%%%%%%%%%%%%%%%%%%%%%%%%%%%%%%%%%%%%%%%%%%%%%%%%%%%%%%%%%%%%%%%%
\section{Conclusion}
%%%%%%%%%%%%%%%%%%%%%%%%%%%%%%%%%%%%%%%%%%%%%%%%%%%%%%%%%%%%%%%%
%%%%%%%%%%%%%%%%%%%%%%%%%%%%%%%%%%%%%%%%%%%%%%%%%%%%%%%%%%%%%%%%
%%%%%%%%%%%%%%%%%%%%%%%%%%%%%%%%%%%%%%%%%%%%%%%%%%%%%%%%%%%%%%%%

We have demonstrated that if the experiments can be performed a sufficient
number of times, then using ``Forced Purity" routine, tomography can
be performed in a quick and robust manner. However, as the entropy
of a state increases, a much more expensive MLE routine has to be
used to perform tomography, which does not scale well as the number
of qubits increases. Quantum computing requires only
pure state tomography, for which we have obtained a scalable and efficient
routine \footnote{The runtimes of the routines mentioned in this
paper can be improved linearly using parallel computation. However, because
the complexity increases exponentially with the number of qubits, an
efficient routine to performing tomography is crucial.}.

\begin{acknowledgments}
The authors would like to thank Robin Blume-Kohout, Ren\'e Stock and
Rob Adamson for stimulating discussions and useful comments.
This work was supported by the U.S. Army Research Office, NSERC and Project
OpenSource.
\end{acknowledgments}

%%%%%%%%%%%%%%%%%%%%%%%%%%%%%%%%%%%%%%%%%%%%%%%%%%%%%%%%%%%%%%%%
%%%%%%%%%%%%%%%%%%%%%%%%%%%%%%%%%%%%%%%%%%%%%%%%%%%%%%%%%%%%%%%%
%%%%%%%%%%%%%%%%%%%%%%%%%%%%%%%%%%%%%%%%%%%%%%%%%%%%%%%%%%%%%%%%
%%%%%%%%%%%%%%%%%%%%%%%%%%%%%%%%%%%%%%%%%%%%%%%%%%%%%%%%%%%%%%%%
%%%%%%%%%%%%%%%%%%%%%%%%%%%%%%%%%%%%%%%%%%%%%%%%%%%%%%%%%%%%%%%%
%%%%%%%%%%%%%%%%%%%%%%%%%%%%%%%%%%%%%%%%%%%%%%%%%%%%%%%%%%%%%%%%
\appendix
%%%%%%%%%%%%%%%%%%%%%%%%%%%%%%%%%%%%%%%%%%%%%%%%%%%%%%%%%%%%%%%%
%%%%%%%%%%%%%%%%%%%%%%%%%%%%%%%%%%%%%%%%%%%%%%%%%%%%%%%%%%%%%%%%
%%%%%%%%%%%%%%%%%%%%%%%%%%%%%%%%%%%%%%%%%%%%%%%%%%%%%%%%%%%%%%%%
\section{Matrix Differential Calculus Theorems\label{sec:Matrix-Differential-Calculus}}
%%%%%%%%%%%%%%%%%%%%%%%%%%%%%%%%%%%%%%%%%%%%%%%%%%%%%%%%%%%%%%%%
%%%%%%%%%%%%%%%%%%%%%%%%%%%%%%%%%%%%%%%%%%%%%%%%%%%%%%%%%%%%%%%%
%%%%%%%%%%%%%%%%%%%%%%%%%%%%%%%%%%%%%%%%%%%%%%%%%%%%%%%%%%%%%%%%

The following theorems were used to derive an analytic expression to the gradient
of the MLE function, which is more computationally efficient than the finite-difference gradient computation as the MLE function itself is expensive to evaluate.
\\

If $M$ is a real-valued matrix and $T=X+i\cdot Y,\, T\in{\cal C}$, then
\cite{Matrix Differential Calculus,Kronecker Products and Matrix Calculus,Matrix Calculus and Zero-One Matrices}:
\begin{equation}
\frac{\partial M}{\partial T}=\frac{\partial M}{\partial X}+i\cdot\frac{\partial M}{\partial Y}.\label{eq:matrix complex derivative rule}\end{equation}
The following are defined for real square matrices~\cite{Kronecker Products and Matrix Calculus}:
\begin{eqnarray}
\frac{\partial Tr\{ X^{T}Y\}}{\partial X}&=&Y,\label{eq:M1}\\
\frac{\partial Tr\{ X^{T}X\}}{\partial X}&=&2X,\label{eq:M2}\\
\frac{\partial Tr\{ KX^{T}Y\}}{\partial X}&=&YK,\label{eq:M3}\\
\frac{\partial Tr\{ KX^{T}X\}}{\partial X}&=&XK+XK^{T}.\label{eq:M4}
\end{eqnarray}
We begin by observing that
\begin{equation}
Tr\{ T^{\dagger}(\vec{t})T(\vec{t})\}=\sum_{\nu=0}^{4^{n}-1}t_{\nu}^{2},\label{eq:Tr T dag T}\end{equation}
which immediately implies that
\[
\frac{\partial Tr\{ T^{\dagger}(\vec{t})T(\vec{t})\}}{\partial t_{\nu}}=2t_{\nu}.\]
Hence, using matrix calculus, we have the result
\begin{equation}
\frac{\partial Tr\{ T^{\dagger}(\vec{t})T(\vec{t})\}}{\partial T}=2X+i\cdot 2Y.\label{eq:Tr T dag T dT}\end{equation}
Eq.~\eqref{eq:Tr T dag T dT} is a compact means of stating the result of Eq.~\eqref{eq:Tr T dag T} using a $2^{n}\times2^{n}$ matrix, where the
value of the derivative is stored in the original position of $t_{\nu}$
in the Cholesky-decomposed matrix $T(\vec{t})$.  This is the general idea behind \emph{all} matrix calculus results
we have used. We could have also obtained the same result by applying
matrix calculus directly. For example, denote
\[
\Phi=T^{\dagger}(\vec{t})T(\vec{t}).\]
Then using Eq.~\eqref{eq:M1} and Eq.~\eqref{eq:M2}
\[
\frac{\partial Tr\{\Phi\}}{\partial X}=2X+i\cdot Y-i\cdot Y=2X\]
and
\[
\frac{\partial Tr\{\Phi\}}{\partial Y}=i\cdot X-i\cdot X+2Y=2Y.\]
This is consistent with our result in Eq.~\eqref{eq:Tr T dag T dT}.\\

Next, we set out to compute $\frac{\partial Tr\{\hat{\Pi}_{\nu}\Phi\}}{\partial T}$.
Recall that $\hat{\Pi}_{\nu}=K_{\nu}+i\cdot\Lambda_{\nu}$, hence,

\begin{eqnarray}
\hat{\Pi}_{\nu}\Phi&=&(K_{\nu}+i\cdot\Lambda_{\nu})(X^{T}X+i\cdot X^{T}Y-i\cdot Y^{T}X+Y^{T}Y)\nonumber\\
&=&K_{\nu}X^{T}X+K_{\nu}Y^{T}Y+\Lambda_{\nu}Y^{T}X-\Lambda_{\nu}X^{T}Y+\nonumber\\
&&i\cdot(K_{\nu}X^{T}Y+\Lambda_{\nu}X^{T}X+\Lambda_{\nu}Y^{T}Y-K_{\nu}Y^{T}X).\nonumber\\
\label{eq:mu*Phi}\end{eqnarray}
Applying Eq.~\eqref{eq:M3} and Eq.~\eqref{eq:M4} to the real part
of Eq.~\eqref{eq:mu*Phi} we obtain
\begin{equation}
\frac{\partial Tr\{\hat{\Pi}_{\nu}\Phi\}}{\partial X}=XK_{\nu}+XK_{\nu}^{T}-Y\Lambda_{\nu}+Y\Lambda_{\nu}^{T}\end{equation}
and
\begin{equation}
\frac{\partial Tr\{\hat{\Pi}_{\nu}\Phi\}}{\partial Y}=X\Lambda_{\nu}-X\Lambda_{\nu}^{T}+YK_{\nu}+YK_{\nu}^{T}.\end{equation}
Observe that $\forall\nu,\,\hat{\Pi}_{\nu}=\hat{\Pi}_{\nu}^{\dagger}$ yields
$\Lambda_{\nu}=-\Lambda_{\nu}^{T}$ and $K_{\nu}=K_{\nu}^{T}$, therefore
\begin{eqnarray}
\frac{\partial Tr\{\hat{\Pi}_{\nu}\Phi\}}{\partial X}&=&2XK_{\nu}-2Y\Lambda_{\nu},\\
\frac{\partial Tr\{\hat{\Pi}_{\nu}\Phi\}}{\partial Y}&=&2X\Lambda_{\nu}+2YK_{\nu}.
\end{eqnarray}
Substituting the above two equations into Eq.~\eqref{eq:matrix complex derivative rule}
we obtain the result described in Sec.~\ref{sub:Matrix-Calculus-Derivation}.

%%%%%%%%%%%%%%%%%%%%%%%%%%%%%%%%%%%%%%%%%%%%%%%%%%%%%%%%%%%

\end{document}